\documentclass[a4paper,fleqn,usenatbib]{mnras}

\usepackage{newtxtext,newtxmath}
\usepackage[utf8]{inputenc}
\usepackage[T1]{fontenc}
\usepackage{ae,aecompl}

\usepackage{graphicx}	
\usepackage{amsmath}	
\usepackage{amssymb}	

\title[Optical excess of dim isolated neutron stars]{OPTICAL EXCESS OF DIM ISOLATED NEUTRON STARS}

\author[\"{U}. Ertan, \c{S}. \c{C}al\i\c{s}kan and M. A. Alpar]{
\"{U}.  Ertan,$^{1}$\thanks{E-mail: unal@sabaciuniv.edu}
\c{S}. \c{C}al\i\c{s}kan$^{1}$ and
M. A. Alpar$^{1}$
\\
$^{1}$Sabanc\i\ University, 34956, Orhanl\i\, Tuzla, \.Istanbul,
Turkey\\
}

\date{Accepted XXX. Received YYY; in original form ZZZ}

\pubyear{2017}

\begin{document}
\label{firstpage}
\pagerange{\pageref{firstpage}--\pageref{lastpage}}
\maketitle

\def\be{\begin{equation}}
\def\ee{\end{equation}}
\def\ba{\begin{eqnarray}}
\def\ea{\end{eqnarray}}
\def\m{\mathrm}
\def\d{\partial}
\def\R{\right}
\def\L{\left}
\def\a{\alpha}
\def\acold{\alpha_\mathrm{cold}}
\def\ahot{\alpha_\mathrm{hot}}
\def\Mdotstar{\dot{M}_\ast}
\def\Omegastar{\Omega_\ast}
\def\Omegadot{\dot{\Omega}}
\def\OmegaK{\Omega_{\mathrm{K}}}
\def\Mdotin{\dot{M}_{\mathrm{in}}}
\def\Mdotd{\dot{M}_{\mathrm{d}}}

\def\Mdotcrit{\dot{M}_{\mathrm{crit}}}
\def\Mdotout{\dot{M}_{\mathrm{out}}}

\def\Mdot{\dot{M}}
\def\Edot{\dot{E}}
\def\Pdot{\dot{P}}
\def\nudot{\dot{\nu}}
\def\Msun{M_{\odot}}
\def\Fopt{F_{\mathrm{opt}}}
\def\Lin{L_{\mathrm{in}}}
\def\Lcool{L_{\mathrm{cool}}}
\def\Rin{R_{\mathrm{in}}}
\def\rin{r_{\mathrm{in}}}
\def\rlc{r_{\mathrm{LC}}}
\def\rout{r_{\mathrm{out}}}
\def\rinmax{r_{\mathrm{in,max}}}
\def\rco{r_{\mathrm{co}}}
\def\re{r_{\mathrm{e}}}
\def\Ldisc{L_{\mathrm{disc}}}
\def\Lx{L_{\mathrm{x}}}
\def\Ld{L_{\mathrm{d}}}
\def\Lopt{L_{\mathrm{opt}}}
\def\Fx{L_{\mathrm{x}}}
\def\Fopt{F_{\mathrm{opt}}}
\def\Md{M_{\mathrm{d}}} 
\def\NH{N_{\mathrm{H}}}
\def\dEb{\delta E_{\mathrm{burst}}}
\def\dEx{\delta E_{\mathrm{x}}}
\def\Bstar{B_\ast}\def\uff{\upsilon_{\mathrm{ff}}}
\def\Bb{\beta_{\mathrm{b}}}
\def\Be{B_{\mathrm{e}}}
\def\Bp{B_{\mathrm{p}}}
\def\Bz{B_{\mathrm{z}}}
\def\Bfi{B_{\mathrm{|phi}}}
\def\BA{B_{\mathrm{A}}}
\def\tint{t_{\mathrm{int}}}
\def\tdiff{t_{\mathrm{diff}}}
\def\r_m{r_{\mathrm{m}}}
\def\rA{r_{\mathrm{A}}}
\def\BA{B_{\mathrm{A}}}
\def\rS{r_{\mathrm{S}}}
\def\rp{r_{\mathrm{p}}}
\def\Tp{T_{\mathrm{p}}}
\def\dMin{\delta M_{\mathrm{in}}}
\def\Rc{\R_{\mathrm{c}}}
\def\Teff{T_{\mathrm{eff}}}
\def\uff{\upsilon_{\mathrm{ff}}}
\def\Tirr{T_{\mathrm{irr}}}
\def\Firr{F_{\mathrm{irr}}}
\def\Tcrit{T_{\mathrm{crit}}}
\def\P0min{P_{0,{\mathrm{min}}}}
\def\Av{A_{\mathrm{V}}}
\def\ah{\alpha_{\mathrm{hot}}}
\def\ac{\alpha_{\mathrm{cold}}}
\def\tc{\tau_{\mathrm{c}}}
\def\p{\propto}
\def\m{\mathrm}
\def\fast{\omega_{\ast}}
\def\Uff{\upsilon_{\mathrm{ff}}}
\def\Ufi{\upsilon_{\fi}}
\def\Ur{\upsilon_{\mathrm{r}}}
\def\UK{\upsilon_{\mathrm{K}}}
\def\Uesc{\upsilon_{\mathrm{esc}}}
\def\Uout{\upsilon_{\mathrm{out}}}
\def\Uphi{\upsilon_{\phi}}
\def\Udiff{\upsilon_{\mathrm{diff}}}
\def\Ure{\upsilon_{\mathrm{r,e}}}
\def\U{\upsilon}
\def\UB{\upsilon_{\mathrm{B}}}
\def\tauB{\tau_{\mathrm{B}}}
\def\hA{h_{\mathrm{A}}}
\def\he{h_{\mathrm{e}}}
\def\cs{c_{\mathrm{s}}}
\def\cse{c_{\mathrm{s,e}}}
\def\hin{h_{\mathrm{in}}}
\def\rhop{\rho^{\prime}}
\def\rhod{\rho_\mathrm{d}}
\def\rhos{\rho_\mathrm{s}}
\def\rhodp{\rho_\mathrm{d}^{\prime}}
\def\rhoe{\rho_\mathrm{e}}
\def\rhoout{\rho_\mathrm{out}}
\def\Alfven{Alfv$\acute{\mathrm{e}}$n~}
\def\418{SGR 0418+5729}
\def\142{AXP 0142+61}
\def\Caliskan{\c{C}al{\i}\c{s}kan~}
\def\ql{\textquotedblleft}
\def\qr{\textquotedblright~}
\def\gpers{g s$^{-1}$}
\def\ergpers{erg s$^{-1}$}
\def\percm2{cm$^{-2}$}
\begin{abstract}

The optical excess in the spectra of dim isolated neutron stars (XDINs) is a significant fraction of their rotational energy loss-rate. This is strikingly different from the situation in isolated radio pulsars. We investigate this problem in the framework of the fallback disc model. The optical spectra can be powered by magnetic stresses on the innermost disc matter, as the energy dissipated is  emitted as blackbody radiation mainly from the inner rim of the disc. In the fallback disc model, XDINs are the sources evolving in the propeller phase with similar torque mechanisms. In this this model, the ratio of the total magnetic work that heats up the inner disc matter is expected to be similar for different XDINs. Optical luminosities that are calculated consistently with the the optical spectra and the theoretical constraints on the inner disc radii give  very similar ratios of the optical luminosity to the rotational energy loss rate for all these sources. These ratios indicate that a significant fraction of the magnetic torque heats up the disc matter while the remaining  fraction expels disc matter from the system. For XDINs, the contribution of heating by X-ray irradiation to the optical luminosity is negligible in comparison with the magnetic heating. The correlation we expect between the optical luminosities and the rotational energy loss-rates of XDINs can be a property of the systems with low X-ray luminosities, in particular those in the propeller phase.

\end{abstract}




\section{Introduction}

Dim isolated neutron stars (XDINs) form one of the young neutron star populations with several distinguishing properties (see e.g. Haberl 2007 for a review). Among other single neutron star systems, they have relatively low X-ray luminosities ($\sim 10^{31} - 10^{32}$ \ergpers) that hinder their detection at large distances. All seven confirmed  XDINs lie within a distance of 500 pc, and have cooling ages less than $10^6$ yr. This indicates that they are rather abundant in the Milky Way with birth rates comparable to those of radio pulsars (Popov, Turolla \& Possenti  2006).  All seven XDIN sources were detected in the optical  band with luminosities higher than the extrapolations of their X-ray spectra, while none of them have confirmed detection in the radio band. Another distinguishing property of  XDINs is their period clustering in the same range ($P \sim 3 - 12$ s ) as that of anomalous X-ray pulsars (AXPs) and soft gamma repeaters (SGRs) (for a review of AXPs and SGRs see Mereghetti 2008).  The source of X-ray luminosity of XDINs is likely to be the intrinsic cooling  of the neutron star with effective temperatures $\sim 50 - 100$ keV (Haberl 2007; Turolla 2009; Kaplan et al. 2011). 

What is the torque mechanism slowing down these systems? A neutron star evolving in vacuum loses rotational energy and angular momentum through magnetic dipole radiation. In this case, the dipole field strength on the pole of the star, $B_0$, can be estimated from the observed period, $P$, and period derivative, $\Pdot$, using the dipole torque formula which gives $B_0 \simeq 6.2 \times 10^{19} (P \Pdot)^{1/2}$. For XDINs, $B_0$ values inferred from the vacuum dipole torque assumption are in the  $10^{13} - 10^{14}$ G range, close to those of AXP and SGRs ($B_0 \gtrsim 10^{14}$ G).  In the presence of a fallback disc around an XDIN, disc torques dominate the dipole torque.  In the fallback disc model (Chatterjee, Hernquist \& Narayan 2000; Alpar 2001),  estimated $B_0$ values can be about one or two orders of magnitude lower than those inferred from the magnetic dipole torque formula. Neutron stars with fallback discs can reach the individual source properties of XDINs and AXP/SGRs   with $B_0 \lesssim 10^{12}$ G  and  $B_0 \gtrsim 10^{12}$ G  respectively (Ertan et al. 2014; Benli \& Ertan 2016).    

In this work, we focus on the optical emission properties of XDINs (Kaplan et al. 2011 and references therein) in the fallback disc model. We will use the results of earlier work on the long-term evolution of individual XDINs (Ertan et al. 2014) and on the inner disc radius of neutron stars in the propeller phase (Ertan 2017). These results indicate that XDINs are evolving in the propeller phase under the effect of disc torques. 
The source of the X-ray luminosity is the intrinsic cooling of the  neutron star. With the current mass-flow rates of XDINs ($\lesssim 10^9$ \gpers) expected from the long-term evolution model, viscous dissipation  in the inner disc cannot produce the optical luminosity of these sources. The X-ray irradiation of the upper and lower surfaces of the disc is not sufficient either. We think of two alternative sources that can power the optical emission of XDINs, namely the X-ray irradiation of the inner rim of the disc and the magnetic stresses that heat up the matter in the disc-field interaction region. For both heating mechanisms, we expect that the surface of the inner rim of the disc is the main site of optical/UV emission. Depending on the current X-ray luminosity, disc mass-flow rate and and the rotational energy loss-rate, one of these mechanisms could dominate the other. Our results exclude X-ray irradiation as a possible source with enough power for the optical excess. We find that the magnetic stresses acting on the inner disc, which is likely to produce a strong relation between the rotational power and the heating rate, could be the source of the optical luminosity of XDINs.    

In Section 2, we estimate the upper and lower limits on the total optical fluxes of XDINs from the available optical data. In Section 3, we investigate the inner disc conditions of XDINs and show that the narrow disc-field interaction boundary heated by the magnetic stresses can produce the observed optical properties of the six XDINs with confirmed period and period derivatives. We discuss our results in Section 4 and summarise our conclusions in Section 5.

\section{Optical spectra and luminosities of XDINs}

The model X-ray spectra seen in Figs. 1--6 were obtained by Kaplan et al. (2011) from the blackbody model fits to the unabsorbed X-ray data of XDINs with effective temperatures given in Table 1. This emission is very likely to be produced by the intrinsic cooling of the neutron stars. All these sources have detections in the optical/UV bands (Kaplan et al. 2011, and references therein). It is seen that the optical data of the sources remain above the extrapolation of the X-ray spectrum to low energies. This is known as the optical excess of XDINs. 

With the current properties of XDINs indicated by the long-term evolution model with fallback discs, heating by viscous dissipation  across  the disc is negligible compared to the X-ray irradiation flux, $\Firr$, provided by the cooling luminosity of the neutron star. At the low X-ray luminosities of XDINs, $\Firr$ is not sufficient to produce the optical excess from the upper and lower surfaces of the disc. For a neutron star with mass $M = 1.4 \Msun$ and radius $R = 1\times 10^6$ cm, the irraditation flux can be written as $\Firr \simeq 1.2 ~C~ \Lx / (\pi r^2)$, where the irradiation efficiency $C \p (h/r)$  includes effects of irradiation geometry and the albedo of the disc surfaces (Fukue 1992). Pressure scale-height $h \p r^{9/8}$, and $h/r \p r^{1/8}$ has a weak $r$ dependence.  During the accretion phase the value of $C$ is estimated to be  $\gtrsim 10^{-4}$ for XDINs for a large range of accretion rates, like in low-mass X-ray binaries (Dubus et al. 1999). After termination of the accretion phase of XDINs, the mass-flow rate of the disc, $\Mdotd$, decreases rapidly to very low levels ($\Mdotd \lesssim 10^9$ \gpers), which also decreases the $h/r$ ratios to less than about $10^{-3}$ and the irradiation efficiency $C$ to a few $10^{-5}$ for  all XDINs. 

All seven XDINs have upper limits in the infrared/\textit{SPITZER} bands (Posselt et al. 2014). We obtain the illustrative model infrared spectra ($\nu < 3\times 10^{14}$ Hz) given in Figs. 1--6 with $C = 3\times 10^{-5}$ and $\cos i = 0.5$ where $i$ is the inclination angle between the normal of the disc and the line of sight of the observer. The inner disc radii taken to obtain the IR spectra consistently match the inner disc radii that can  produce the optical spectra (Section 3).  As seen in Figs. 1--6, the contribution of the emission from the irradiated upper/lower disc surfaces to the optical luminosity is too weak  to account for the observed optical spectra. The detections of two sources, namely RX J0806.4-4123 and RX J2143.0+0654,  in the \textit{Herschel} PACS red band should be taken with some caution, because whether these detections are from the sources themselves or from their bright neighbours is not clear yet, due to the so-called ``confusing neighbours" (Posselt et al. 2014).

In this work, we propose that the optical flux is emitted mainly from the inner rim of the disc in the form of a blackbody spectrum. To test this idea and the alternative sources for powering the optical luminosity, we first estimate the total observed optical flux $\Fopt$ and the effective temperature, $\Teff$, of the emitting region from the spectra of the sources.   
The two model curves (dashed and dashed-dotted) seen in Figs. 1--6 correspond to the minimum and the  maximum $\Teff$ allowed by the error bars of the data points respectively. The $\Fopt$ corresponding to a particular $\Teff$ is found by integrating the  spectrum produced by this $\Teff$. For each XDIN source, $\Fopt$ and $\Teff$ values are given in Table 1. In Section 3, we use these results to estimate the optical luminosities of the  sources in consistence with their inner disc properties.  

\begin{table}
\caption{\label{table:opticalinfo} The optical and X-ray fluxes and corresponding blackbody temperatures of the six XDINs. The full names of the sources are: RX J1856.5-3754, RX J0720.4-3125, RX J2143.0+0654, RX J1308.6+2127, RX J0806.4-4123, RX J0420.0-5022. The X-ray temperatures are taken from Kaplan et al. (2011).}
\begin{minipage}{\linewidth}
\begin{center}
\begin{tabular}{c|c|c|c|c} \hline \hline
 Source & $F_{\mathrm{X}}$ &$F_{\mathrm{opt}}$ & $kT_{\mathrm{X}}$&$k\Teff$ \\  
 &($10^{-12}$ erg/s/cm$^2$)&( $10^{-14}$erg/s/cm$^2$)& (eV)&(eV)\\ 
 \hline
1856 & 16.6 & 8.1$-$24 & 63.5 &4.3$-$6.9\\
0720& 9.0 & 1.9$-$39& 88.4& 4.1$-$12.9\\
2143& 2.3 & 0.28$-$0.31&104 &2.0$-$2.3\\
1308& 3.6& 0.20$-$2.6&102&  3.3$-$8.6 \\
0806&3.2 &0.25$-$8.3&87.2&3.4$-$12.9\\
0420&0.51&3.8$-$7.3&45&8.6$-$11.2\\ 
\hline \hline
\end{tabular}\\
\end{center}
\end{minipage}
\end{table}

\begin{figure}
\includegraphics[height=.35\textheight,angle=270]{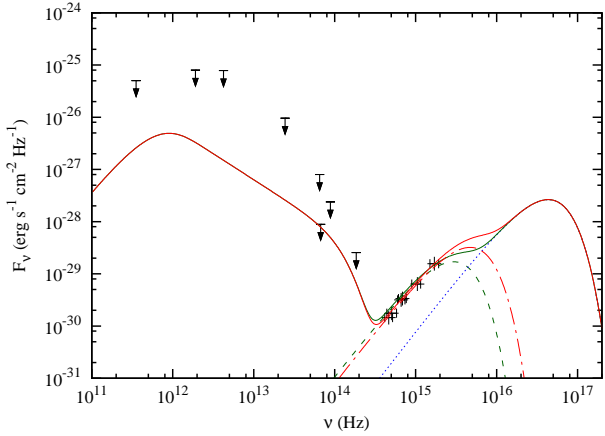}
\caption{The spectrum of RX J1856.5-3754. The arrows are infrared/\textit{SPITZER} upper limits (Posselt et al. 2014) and the black data points are the optical/UV detections (Kaplan et al. 2011). The dashed (dark green) and dot-dashed (red) model blackbody spectra are obtained with the minimum and the maximum $\Teff$ that can fit the optical/UV data. The blackbody spectrum in X-rays seen in the figure (dotted blue curve) is obtained with the best-fitting parameters from Kaplan et al. (2011). The IR part of the model spectrum ($\nu < 3 \times 10^{14}$ Hz) is the disc-blackbody spectrum expected from the extended surface of the disc irradiated by the cooling luminosity of RX J1856.5-3754 with $\cos i = 0.5$ and $C= 3\times 10^{-5}$. The solid (red) curve represent the total model spectrum from IR to X-rays. The model parameters are given in Table 1 
}
\end{figure}

\begin{figure}

\includegraphics[height=.35\textheight,angle=270]{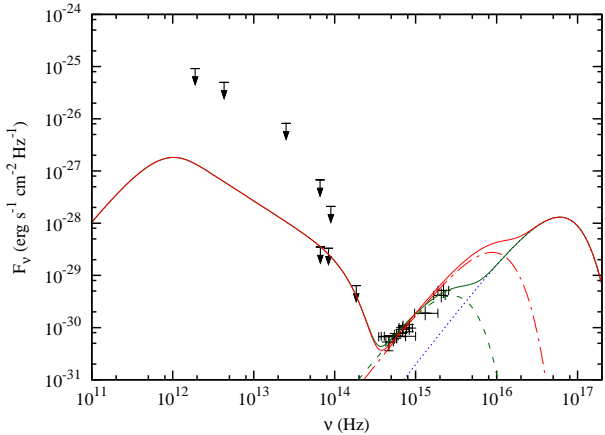}
\caption{The spectrum of RX J0720.4-3125. Model parameters are given in Table 1. }
\end{figure}

\begin{figure}
\includegraphics[height=.35\textheight,angle=270]{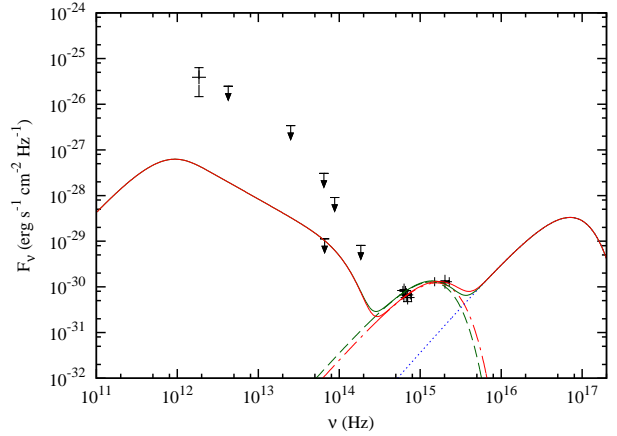}
\caption{The spectrum of RX J2143.0+0654. There are three different $B$ band detections of this source (Kaplan et al. 2011, Zane et al. 2008 and Schwope et al. 2009). The average of these values was used for the rim blackbody fits. The relatively low $R$ band flux (i.e. the small slope between the two optical detections) restricts the peak location of the optical blackbody spectrum and $\Teff$. The leftmost data point is the so-called "confusing neighbor", which could be emission from a nearby bright star instead of RX J2143.0+0654 (Posselt et al. 2014). Model parameters are given in Table 1.}
\end{figure}

\begin{figure}
\includegraphics[height=.35\textheight,angle=270]{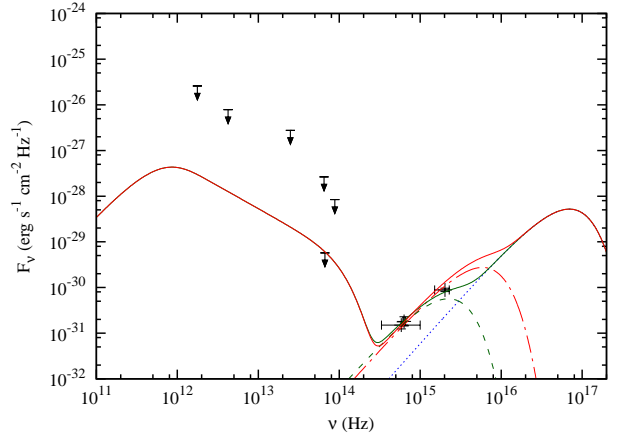}
\caption{The spectrum of RX J1308.6+2127. The disc blackbody spectrum was obtained with $\cos i = 0.4$ . Model parameters are given in Table 1.}
\end{figure}

\begin{figure}
\includegraphics[height=.35\textheight,angle=270]{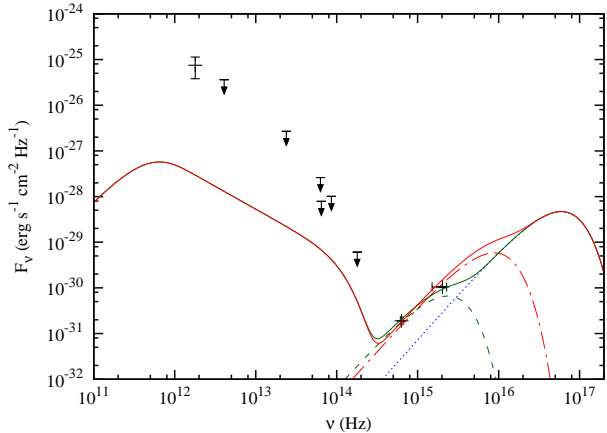}
\caption{The spectrum of RX J0806.4-4123. Model parameters are given in Table 1. The leftmost data point is the so-called "confusing neighbour", which may be from a nearby bright star (Posselt et al. 2014). }
\end{figure}

\begin{figure}
\includegraphics[height=.35\textheight,angle=270]{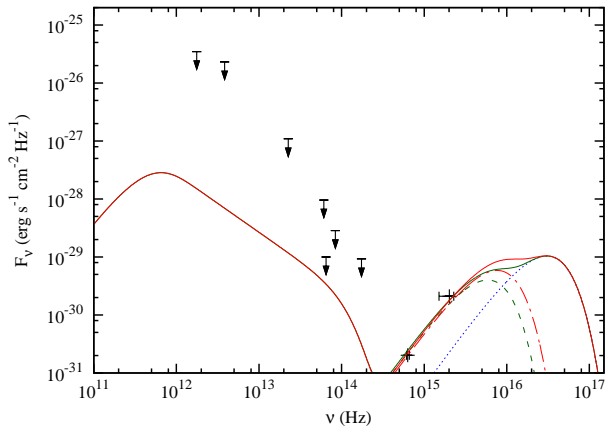}
\caption{The spectrum of RX J0420.0-5022. Model parameters are given in Table 1. }
\end{figure}

\section{What powers the optical emission of XDINs?}

For isotropic X-ray emission from the star the fraction of the X-ray luminosity, $\Lx$,  that is absorbed by the disc's inner rim is estimated to be $\sim \hin / \rin$ where $\hin$ is the half thickness of the disc at the inner radius of the disc, $\rin$.   The $\hin / \rin$ ratio depends very weakly on both the disc mass-flow rate, $\Mdotd$, and $\rin$. We expect the ratio of the optical luminosity, $\Lopt$, to $\Lx$ to be similar for different XDINs, if the inner disc is heated mainly by the X-ray irradiation.   The second alternative source to power the optical emission is the interaction between the inner disc and the dipole field of the star.  Part of the work done by the magnetic stresses, at the expense of the rotational energy of the star, could be converted into heat in the disc. 
Results of recent analytical and numerical work indicate that the inner disc-field interaction takes place in a narrow boundary (Lovelace, Romanova \& Bisnovatyi-Kogan 1995; Hayashi, Shibata \& Matsumoto 1996; Goodson, Winglee \& Boehm 1997; Miller \& Stone 1997; Lovelace, Romanova \& Bisnovatyi-Kogan 1999; Uzdensky, K\"{o}nigl \& Litwin 2002; Uzdensky 2004; Ustyugova et al. 2006). This means that, like X-ray irradiation, the magnetic stresses also heat up the innermost region of the disc, and the resultant radiation is likely to be emitted mostly from the surface of the  inner rim of the disc.  Independently of the  details of the heating inside the boundary, if this mechanism, rather than the X-ray irradiation, is responsible for the optical emission, the ratio of $\Lopt$ to the rotational energy loss-rate, $\Edot$, is expected to be similar for different XDINs that slow down with similar propeller torques. The rotational  power $ \Edot= I \Omega |\Omegadot|$ of XDINs, where $I$ is the moment of inertia, $ \Omega$ is the rotational frequency of the neutron star, and $\Omegadot$ is the spin-down rate, is given in Table 2, along with the periods, period derivatives, estimated distances, and the X-ray luminosities of the sources. To estimate $\Lopt$ using $\Teff$ obtained from the spectral fits, we need the inner disc radius $\rin$, and the area, $A$, of the inner rim of the disc.  

\begin{table}
\caption{\label{table:table2_ppdotinfo} The period, period derivative, distance, luminosity and $\Edot = I \Omega \Omegadot$ values of six XDINs. The X-ray luminosity values are taken from Ertan et al (2014). The distances are taken from the references: (1) Posselt et al. 2007, (2) Kaplan, van Kerkwijk \& Anderson 2007, (3) Hambaryan et al. 2011.}
\begin{minipage}{\linewidth}
\begin{center}
\begin{tabular}{c|c|c|c|c|c|c} \hline \hline
   & 1856 & 0720 & 2143 & 1308 & 0806 & 0420 \\ \hline
$P$ (s) & 7.055 & 8.39 & 9.44 & 10.31 & 11.37 & 3.45\\
$\Pdot$ (10$^{-13}$ s/s) & 0.297 & 0.698 & 0.4 & 1.120 & 0.55 & 0.28\\
$d$ (pc) & 135$^{1}$ & 270$^{1,2}$ & 400$^{1}$ & 380$^{3}$ & 240$^{1}$ & 345$^{1}$\\
$\Lx$~(10$^{31}$ \ergpers) & 9.5  & 16 & 11 & 7.9 & 2.5 & 2.6\\
$\Edot$~(10$^{30}$ \ergpers) & 3.3 & 4.7 & 1.9 & 4.0 & 1.5 & 27\\
\hline \hline
\end{tabular}\\
\end{center}
\end{minipage}
\end{table}

\subsection{The inner disc radius in the propeller phase}

The XDINs are expected to be in the propeller phase of their evolution in interaction with a fallback disc (Ertan et al. 2014). 
Where is the inner disc radius in the propeller phase? At the co-rotation radius $\rco = (GM)^{1/3} \Omega^{-2/3}$, where $G$ is the gravitational constant, and $M$ is the mass of the neutron star, the field lines and the disc matter move with the same angular velocity. Since the escape speed $\Uesc = \sqrt{2} ~r  \OmegaK$ where $\OmegaK$ is the local Keplerian angular velocity in the disc,  the disc matter  can be accelerated to above $\Uesc$ at radii $r$ where $\OmegaK(r) / \Omega \geq \sqrt{2}$. Since $\OmegaK \p r^{-3/2}$, the field lines co-rotating with the star have speeds greater than $\Uesc(r)$ at radii greater than 
\be
r_1 = 2^{1/3} \rco = 1.26 ~\rco. 
\label{10}
\ee 
That is, the inner disc radius $\rin$ should be greater than $r_1 = 1.26 ~\rco$ in a steady propeller phase. 

Recently, from simple analytical calculations, Ertan (2017) estimated the maximum possible inner disc radius in the propeller phase.
Theoretical and numerical work on the disc-field interaction show that the field lines interact with the inner disc in a narrow boundary (see e.g. Lovelace et al. 1995). In this interaction region, the field lines cannot slip through the disc in the azimuthal direction due to the long magnetic diffusion time-scale that is similar to the viscous time-scale  (Fromang \& Stone 2009), and is still orders of magnitude smaller than the time-scale of interaction between the field lines and the inner disc matter, $\tint \simeq |\Omega - \OmegaK|^{-1}$. The field lines inflate and open up within $\tint$, expel the matter along the open field lines, and subsequently reconnect on the dynamical time-scale completing the cycle (Lovelace et al. 1999; Ustyugova et al. 2006). Outside the boundary, the disc and the field lines are expected to be decoupled. The outflow speed of the matter depends on the dipole field strength and the mass density of the disc matter. At a given radius, the maximum amount of angular momentum that can be injected into the matter is limited by the interaction time-scale, unlike in models assuming that the closed field lines could remain threaded across a large boundary region. For a given mass-flow rate of the disc and dipole field strength, the minimum angular momentum transfer required to accelerate the matter to above the escape speed can be sustained up to a certain critical radius. This is the maximum inner disc radius at which a steady propeller mechanism can be built up, and is given by   
\be
\rin \simeq 3.8 \times 10^{-2} ~\a_{-1}^{2/7} ~(\omega - 1)^{-4/7}~ \left(\frac{\hin}{\rin}\right)_{-2}^{4/7} ~\rA
\label{11}
\ee  
(Ertan 2017)
where $\a_{-1} = \a / 0.1$ is the kinematic viscosity parameter (Shakura \& Sunyaev 1973), $(\hin/\rin)_{-2} = (\hin/\rin)/ 10^{-2}$, $\omega = \Omegastar / \OmegaK \geq \sqrt{2}$ is the fastness parameter, and $\rA$ is the conventional \Alfven radius 
\be
\rA \simeq (G M)^{-1/7}~\mu^{4/7}~ \Mdot^{-2/7}         
\label{12}
\ee
(Lamb, Pethick \& Pines 1973; Davidson \& Ostriker 1973) where $\Mdot$ is the spherical accretion rate and $\mu = (B_0 /2)  R^3$ is the magnetic dipole moment of the star. Equation (\ref{11}) is valid only in the propeller phase, that is, when $\omega > \sqrt{2}$. Because of sharp radial dependence of magnetic stresses we expect that the actual inner disc radius is near the radius given by equation (\ref{11}). 
The $\hin / \rin$ ratio should be calculated for an unperturbed geometrically thin disc.
For a standard thin disc, $h/r \p \Mdotd^{3/20} ~r^{1/8}$ has a very weak dependence on both $r$ and the disc mass-flow rate $\Mdotd$ ~(see e.g. Frank, King \& Raine 2002). In equation (\ref{11}), the $h/r$ ratio at the inner disc radius should also be calculated for an unperturbed standard disc (Ertan 2017).

For disc accretion, $\rin$ is usually assumed to be close to the radius where the viscous stresses in a standard disc is balanced  by the magnetic stresses, and with different assumptions, $\rin$ is usually inferred to be close to $\rA$ within a factor of $\sim 2$ (e.g. Frank et al. 2002). It was shown that a steady propeller mechanism cannot be sustained at a radius close to $\rA$ (Ertan 2017). From equation (\ref{11}), it can be calculated that $\rin$ should be at least $\sim 15$ times smaller than $\rA$, in the propeller phase. 

When $\rin$ is sufficiently greater than $\rco$ such that $\omega -1 \sim \omega$, equation (\ref{11}) gives $\rin \p \rA^{7/13} \p \Mdotd ^{-2/13}$. This indicates that once a system enters the propeller phase, $\rin$ remains close to $\rco$ even with very low $\Mdotd$. For an order of magnitude decrease in $\Mdotd$ below the accretion-propeller transition level, the inner disc radius increases only by a factor of 1.4. In the propeller phase, $\rin$ is the radius where all the inflowing matter, with the rate $\Mdotd$, is thrown out of the system with speeds greater than the escape speed. 

The critical accretion rate for the accretion-propeller transition can be estimated by setting $\omega = \sqrt{2}$ in equation (\ref{11}) which gives  
\be
\Mdotcrit \simeq 4 \times 10^{11} ~g s^{-1} ~\a_{-1} ~P^{-7/3} ~ \mu_{30}^2 \left(\frac{\hin}{\rin}\right)_{-3}^{2}
\label{13}
\ee  
(Ertan 2017) where $\mu_{30}$ is the magnetic dipole moment of the star in units of $10^{30}$ G cm$^3$, $P$ is the spin period of the star, and $(\hin / \rin)_{-3}$ is the $h/r$ ratio at the inner disc radius in units of $10^{-3}$, a typical value for XDINs. Above this critical accretion rate, the system is expected to be in the accretion phase. The accretion-propeller transition condition estimated from equation (\ref{13}) is consistent with the observed properties of the transitional milli-second pulsars which show transitions between the accretion powered X-ray pulsar state and the radio pulsar state (Papitto et al. 2015, Archibald et al. 2015). 
From our earlier work on the long-term evolution of XDINs, we have estimated $\mu_{30}$ to be in the $0.2 - 0.7$ range, and the current mass-flow rates  $\Mdotd \lesssim 10^9$  \gpers . 
For all XDINs,  $\Mdotcrit \sim 10^9 - 10^{10}$ \gpers, and the actual disc accretion rates, $\Mdotd$,  are below the critical rate, thus the XDINs should be in the propeller phase currently.    
Using the $\Mdotd$ values estimated from the long-term evolution model for each source, the maximum inner disc radii from equation (\ref{11}) are found to be less than around $2.8 ~\rco$. These results constrain $\rin$ of these sources into a narrow range ($1.26 ~\rco < \rin < 2.8  ~\rco$).  The  $\hin /\rin$ values are weakly dependent on $\Mdotd$, and we find $\hin /\rin \lesssim 10^{-3}$ for all these sources with $\rin \sim \rco$.
These results allow us to constrain  $\rin$ and the area $A = 2\pi \rin 2\hin$ of the rim of the inner disc for a given XDIN source. We will estimate the optical luminosities using these well constrained $\rin$ and $\hin /\rin$ values and the effective temperatures obtained from the spectral fits.   

\subsection{Estimation of the optical luminosity}

To calculate the optical luminosity, $\Lopt$, we need only the total area of the inner rim of the disc, $A$, from the theoretical calculations, and the effective temperature from the observed spectrum.
We define a geometrical factor, $\eta$, that represents the fraction of the total area of the inner rim of the disc, $A$, that is visible and projected onto a plane perpendicular to the line of sight, $A^\prime = \eta A $. Since we can see less than half of the total area A, $\eta$ is estimated to be less than about $ \sin i / \pi$ where $i$ is the angle between the normal of the disc and the line of sight. The total optical luminosity can be written as 
\be
\Lopt = 4 \pi d^2 \eta^{-1} \Fopt ~=~ \sigma \Teff^4 A 
\label{14}
\ee  
where $d$ is the distance to the source, $\sigma$ is the Stefan-Boltzmann constant and $\Teff$ is the effective temperature of the disc's inner rim. In equation (\ref{14}), the inner rim area $A$ of a source is well restricted independently of the spectrum as discussed in Section 3.1. $\Fopt$ is the observed flux obtained from the spectrum with this particular $\Teff$. In the spectral fits, error bars of data points limits $\Teff$ into ranges with different widths for different sources (Table 1). 

For a given source, these restrictions put upper and lower limits on $\Lopt$, and thus on $d^2 / \eta$ value. The maximum $\Lopt$ corresponds either to $\Lopt < \Edot$ or to a lower luminosity calculated by using  with the maximum allowed values of $\Teff$ and $\rin$. The minimum $\Lopt$ is obtained with  $\rin = 1. 26 ~\rco$ and the minimum $\Teff$ permitted by the spectral fits.  The results of our earlier work on the long-term evolution of these sources indicate that they all slow down in the propeller phase with similar torque mechanisms (Ertan et al. 2014). In this model, we expect the fraction of the rotational energy-loss rate of the star heating up the inner disc matter to be  similar for these systems. To test whether the observed optical luminosities of XDINs could be produced by this heating mechanism, we calculate the allowed $\Lopt$ ranges of the six XDINs, and compare their $\Lopt/\Edot$ ratios.

For the illustrative results given in Table 3, it is seen that the $\Lopt / \Edot$ ratios of the six XDINs are found to be similar (0.45 - 0.90). The last term in equation (\ref{14}) does not depend on $\eta$ or $d$. $\Lopt$ is calculated using the allowed values of $A$ from the theoretical model, and $\Teff$ independently from the spectrum. For a given source, calculated $\Lopt$ corresponds to a certain value of $d^2 / \eta$ (see equation (\ref{14})). We have estimated the $\eta$ values for the 6 XDINs using the distances employed earlier in their long-term evolution model. In Table 3, it is seen that these $\eta$ values are distributed in a plausible range ($0.01 \lesssim \eta  \lesssim 0.07$). If the distance of a source is modified, the same $\Lopt / \Edot$ ratio can be obtained by modifying the $\eta$ as well. The maximum value of $\eta$ could be $\sim 0.3$ due to the restrictions by the viewing geometry. This does not impose a significant limitation on distances, and our results for $\Lopt / \Edot$ ratios do not change within a factor  $\gtrsim 2$ of the distances given in Table 2.

In this picture, a self-consistent explanation of the optical excess of XDINs requires similar $\Lopt / \Edot$ ratios for all these sources, and our results show that  this is possible only with high $\Lopt / \Edot$ values close to unity (see Table 3), which indicates that 
a significant fraction of the work done by the magnetic torques is converted into thermal energy in the boundary layer. The resultant blackbody emission from the surface of the inner rim of the disc is in good agreement with the optical data. Note that optical luminosities of XDINs, even assuming isotropic emission, are too high to have  a magnetospheric origin (see Kaplan et al. 2011 for a detailed discussion on the alternative sources of the optical excess for the neutron stars without a fallback disc).     
 
The $\Lopt/\Lx$ ratios  are also given in Table 2.  For RX J0420.0-5022, the ratio is 2.8, which cannot be attained by the X-ray irradiation, since only  a small fraction, $\hin/\rin$, of $\Lx$ can illuminate the inner rim of the disc. For the other sources, the $\Lopt/\Lx$ ratios are greater than $10^{-2}$. These ratios would be possible if $\hin/\rin > 10^{-2}$ which is not likely to be attained with the temperatures that can be sustained with the X-ray irradiation flux illuminating the inner discs of XDINs. More importantly, the $\Teff$ value that can be produced by X-ray irradiation, at the inner rim of the disc ($\sim$ 1 eV) remains much below the minimum $\Teff$ allowed by the optical spectra for all these sources (given in Table 1).
In sum, the X-ray irradiation cannot produce the optical luminosity of XDINs consistently with their optical  spectra. The contribution of X-ray irradiation to the heating of the inner rim of the disc seems to be negligible for XDINs. For other systems with relatively high X-ray luminosities, the X-ray irradiation could dominate the magnetic heating. We estimate that the inner rim of the disc is heated dominantly by the magnetic stresses for the systems with $\Edot / \Lx \gtrsim \hin/\rin$.

\begin{table}
\caption{\label{table:table3_results} The results.}
\begin{minipage}{\linewidth}
\begin{center}
\begin{tabular}{c|c|c|c|c|c|c} \hline \hline
   & 1856 & 0720 & 2143 & 1308 & 0806 & 0420 \\ \hline
$\Lopt $ ($10^{30}$ \ergpers) & 2.5 & 2.8 & 0.84 & 1.6 & 1.4 & 24.1\\
$\Lopt/\Edot$ & 0.75 & 0.59 & 0.45 & 0.57& 0.92 & 0.74\\
$\Lopt/\Lx$ & 0.07 & 0.04& 0.02 & 0.04& 0.07 & 2.8\\
$\rco$~(10$^{8}$ cm) & 6.2  & 6.9 & 7.5 & 7.9 & 8.7 & 3.8\\
$\rin/\rco$ & 1.51  & 1.45 & 2.50 & 1.80 & 1.30 & 1.46\\
$k\Teff$ (eV)  & 4.3  & 4.3 & 2.3 & 3.3 & 3.4 & 8.6\\
$\eta$ & 0.07 &0.06 & 0.07 & 0.02 & 0.01 & 0.03\\
\hline \hline
\end{tabular}\\
\end{center}
\end{minipage}
\end{table}

\section{Conclusions}

The optical luminosities, $\Lopt$, of the six XDINs are orders of magnitude greater than the total optical luminosity that can be produced by the viscous dissipation in their discs. We have estimated and compared $\Lopt$ with possible sources of power from the neutron star. First, we have tested the effect of X-ray irradiation of the disc by the cooling luminosity of the neutron star. For one XDIN, namely  RX J0420.0-5022, $\Lopt > \Lx$ which directly eliminates X-ray irradiation as an alternative source of optical excess of this particular source. For the other XDINs, we have found that that the X-ray irradiation of the upper/lower surfaces and the inner rim of the disc cannot produce the optical luminosities consistently with the effective temperatures indicated by the optical spectra. 

Next, we have considered the rotational energy loss-rate  of the neutron star, $\Edot = I \Omega |\Omegadot|$, which is determined mainly by interaction of the inner disc with magnetosphere of the star in the fallback disc model.  The rotational power $\Edot$ is a measured quantity, and the thermal radiation powered by the heating effect of the magnetic stresses on the disc matter is expected to be emitted mostly from the inner rim of the disc. This interaction mechanism is likely to be similar in different XDIN sources evolving in the propeller phase producing similar $\Lopt / \Edot$  ratios if the source of their $\Lopt$ are indeed powered by the magnetic torques. We have obtained our results with this basic assumption without addressing the details of the magnetic torque and heating mechanisms.
We have estimated the ranges of $\Lopt$ values allowed by the observed spectra, and the constraints on the inner disc radii and thickness placed by the earlier work on the evolution of XDINs and on the inner disc radii in the propeller phase. We have found that the $\Lopt$ values compatible with these constraints
match a similar fraction of $\Edot$ for all the six XDINs (Table 3).

Note that the optical luminosities of XDINs cannot be produced by magnetospheric emission as in isolated neutron stars without fallback discs. Typical $\Lopt/\Edot$ ratio is less than $10^{-6}$ for radio pulsars (Zavlin \& Pavlov 2004) which is several orders of magnitude lower than those in XDINs. 
 
The rotational powers of XDINs are lower than their X-ray luminosities. Nevertheless, our results indicate that a large fraction of their rotational power heats up the inner disc matter, while a small fraction of the X-ray luminosity ($\sim \hin/\rin$) can illuminate the inner rim of the disc.  We have estimated that the inner discs of the systems with  $\Edot / \Lx >  \hin/\rin$ are heated mainly by the magnetic stresses. All the known XDINs show this property. 

To sum up, the optical excess of XDINs can be explained by the emission from the inner rims of their fallback discs due to heating during the propeller process powered by magnetic stresses at the expense of the star's rotational energy. The required neutron star and disc properties are consistent with the results of the long-term evolution model that can reproduce the X-ray luminosity and the rotational properties of the individual sources (Ertan et al. 2014).


\section*{Acknowledgements}

 We acknowledge research support from
T\"{U}B{\.I}TAK (The Scientific and Technological Research Council of
Turkey) through grant 113F354  and from Sabanc\i\ University. MAA is a member of the Science Academy - Bilim Akademisi, Istanbul, Turkey.   












\bsp	
\label{lastpage}
\end{document}